\documentclass[sigconf]{acmart}

\usepackage{xcolor}
\usepackage{lineno}
\usepackage{enumitem}
\usepackage{placeins}
\usepackage{balance}

\begin{document}

\title{From Newborn to Impact: Bias-Aware Citation Prediction}

\author{Mingfei Lu}
\affiliation{%
  \institution{University of Technology Sydney}
  \city{Sydney}
  \country{Australia}
  }
\email{mingfei.lu@student.uts.edu.au}

\author{Mengjia Wu}
\affiliation{%
  \institution{University of Technology Sydney}
  \city{Sydney}
  \country{Australia}
  }
\email{mengjia.wu@uts.edu.au}

\author{Jiawei Xu}
\affiliation{%
  \institution{University of Texas at Austin}
  \city{Austin}
  \country{United States}
  }
\email{jiaweixu@utexas.edu}

\author{Weikai Li}
\affiliation{%
  \institution{University of California, Los Angeles}
  \city{Los Angeles}
  \country{United States}
}
\email{weikaili@cs.ucla.edu}

\author{Feng Liu}
\affiliation{%
  \institution{The University of Melbourne}
  \city{Melbourne}
  \country{Australia}
  }
\email{feng.liu1@unimelb.edu.au}

\author{Ying Ding}
\affiliation{%
  \institution{University of Texas at Austin}
  \city{Austin}
  \country{United States}
}
\email{ying.ding@austin.utexas.edu}

\author{Yizhou Sun}
\affiliation{%
  \institution{University of California, Los Angeles}
  \city{Los Angeles}
  \country{United States}
  }
\email{yzsun@cs.ucla.edu}

\author{Jie Lu}
\affiliation{%
  \institution{University of Technology Sydney}
  \city{Sydney}
  \country{Australia}
}
\email{jie.lu@uts.edu.au}

\author{Yi Zhang}
\authornote{Corresponding author.}
\affiliation{%
  \institution{University of Technology Sydney}
  \city{Sydney}
  \country{Australia}
}
\email{yi.zhang@uts.edu.au}

\renewcommand{\shortauthors}{Mingfei Lu et al.}

\begin{abstract}
As a key to accessing research impact, citation dynamics underpins research evaluation, scholarly recommendation, and the study of knowledge diffusion. Citation prediction is particularly critical for newborn papers, where early assessment must be performed without citation signals and under highly long-tailed distributions. We identify two key research gaps: (i) insufficient modeling of implicit factors of scientific impact, leading to reliance on coarse proxies; and (ii) a lack of bias-aware learning that can deliver stable predictions on lowly cited papers. We address these gaps by proposing a Bias-Aware Citation Prediction Framework, which combines multi-agent feature extraction with robust graph representation learning. First, a multi-agent × graph co-learning module derives fine-grained, interpretable signals, such as reproducibility, collaboration network, and text quality, from metadata and external resources, and fuses them with heterogeneous-network embeddings to provide rich supervision even in the absence of early citation signals. Second, we incorporate a set of robust mechanisms: a two-stage forward process that routes explicit factors through an intermediate exposure estimate, GroupDRO to optimize worst-case group risk across environments, and a regularization head that performs what-if analyses on controllable factors under monotonicity and smoothness constraints. Comprehensive experiments on two real-world datasets demonstrate the effectiveness of our proposed model. Specifically, our model achieves around a 13\% reduction in error metrics (MALE and RMSLE) and a notable 5.5\% improvement in the ranking metric (NDCG) over the baseline methods. The code can be found at https://github.com/Maekfei/BA-Cite.
\end{abstract}




\maketitle

\section{Introduction}
Citation dynamics, as a key to accessing research impact, is crucial for research evaluation, scholarly recommendation, and the study of knowledge diffusion. 
Citation prediction has thus become a particularly significant task for identifying scientific innovation from newborn papers. However, citation data are highly biased. Implicit factors such as reproducibility are not well reflected in modeling, while overly correlated explicit factors lead to shortcuts, e.g., top venue = high citations ~\citep{joachims2017unbiased,schnabel2016recommendations}. Thus, accurately and robustly predicting citations in such a cold-start scenario remains a pressing challenge.
As citation behaviors unfold across large-scale web-based scholarly platforms, addressing this challenge also contributes to robust and generalizable web mining of academic content and information diffusion.

Previous work on citation prediction falls largely into two categories.
(1) Early-cascade models use initial citation dynamics as predictors. DeepCas~\citep{li2017deepcas} treats the citation cascade of a paper as a sequence, generating diffusion pathways through random walks and modeling them with BiGRU and attention to capture early spread signals. SI-HDGNN~\citep{xu2022heterogeneous} further embeds these processes in heterogeneous dynamic academic graphs, combining multi-relational structures with early citation sequences to forecast long-term scientific impact. However, these models require years of waiting for early citations to accumulate, making them ineffective in the cold-start stage when timely prediction is most needed.
(2) Metadata-driven models leverage the author, institution, venue, and related descriptors. For example, HINTS~\citep{10.1145/3442381.3450107} models dynamic heterogeneous information networks, via graph neural networks (GNNs) to capture temporal evolution in citation time series. Cluster-Aware Text-Enhanced HGNN~\citep{yang2023revisiting} integrates signals with cluster-level and textual features to improve prediction. However, citation distributions are highly long-tailed~\citep{10.1145/3442381.3450107,newman2005power}, and these models perform poorly on lowly cited papers. Moreover, they tend to overfit correlations with explicit factors while overlooking deeper implicit factors that may have comparably higher potential to influence citation behaviors in real-world scientific activities, leading to significant performance degradation under distribution shifts.
 
Despite some promising solutions, existing methods cannot achieve accurate and robust citation prediction for particularly lowly cited papers due to the following two research gaps:
\textbf{Gap 1: Insufficient attention to implicit factors, resulting in reliance on strong correlations with explicit factors and poor generalization across domains.}
Current models rely heavily on factors such as author reputation and venue prestige, which correlate with citations but cannot comprehensively cover all decisive determinants of citation behaviors. Lacking fine-grained representations of implicit factors such as topic hotness, reproducibility, and collaboration structure, models would default to superficial correlations, which ultimately degrade performance when the data distribution shifts.
\textbf{Gap 2: A lack of bias-aware models that remain robust on lowly cited papers.}
In this work, bias refers to the group-level prediction disparity induced jointly by the long-tailed citation distribution, feature sparsity in cold-start settings, and the empirical risk minimization (ERM) objective that minimizes average risk~\citep{liu2021just,hashimoto2018fairness}. This structural bias causes models to systematically underperform on low-citation subgroups. 
Most existing methods overlook the issue, since minimizing overall error inherently biases training toward highly cited papers that dominate the loss. As a result, lowly cited papers remain underrepresented and their citation dynamics is poorly predicted, undermining the early evaluation of underrepresented elements in the research community, e.g., early career researchers and emerging research directions.
Based on these gaps, we pose our core question: \begin{center}
\textbf{How can we design bias-aware models that deliver stable predictions on low-citation papers while revealing the underlying drivers of scientific impact?}\end{center}

To bridge these two significant gaps, we propose a Bias-Aware Citation Prediction Framework, termed \textit{BA-Cite}, which combines fine-grained feature extraction with robust GNN learning.
Specifically, \textbf{to tackle Gap 1}, we design a multi-source informed graph learning framework that jointly models agent-derived implicit factors and heterogeneous graph structures. Six agents automatically extract fine-grained features, including reproducibility, text quality, collaboration network, topical hotness, venue prestige and role-aware author reputation, from metadata and external resources. Instead of serving as isolated inputs, these signals are fused with graph-based paper embeddings and propagated through a two-stage predictor, allowing the model to integrate implicit factors with graph context. In this way, the framework reduces reliance on explicit proxies and achieves stronger generalization.  
\textbf{To overcome Gap 2}, we center learning on a two-stage predictor: The model initially estimates an intermediate exposure variable from graph embeddings and agent-derived features, and then predicts citations based on both the exposure estimate and the remaining features, while excluding superficial correlates from direct inputs. Robust learning objectives are attached to the second stage’s outputs: (i) Group Distributionally Robust Optimization (GroupDRO) minimizes the worst-group risk on the prediction loss, countering head-dominated bias; and (ii) a regularization module performs what-if interventions on controllable factors, recomputing exposure and predictions while enforcing monotonicity and smoothness constraints. These objectives are jointly optimized and back-propagated through both stages and the graph encoder, shaping the pipeline toward bias-resistant and consistent behavior.
We conduct systematic evaluations on two large-scale academic datasets, AMiner and OpenAlex. Experimental results demonstrate that, compared with the state-of-the-art baselines, our framework chieves around a 13\% reduction in error metrics (MALE and RMSLE) and a notable 5.5\% improvement in the ranking metric (NDCG).  

The main contributions of this paper are highlighted as follows:  
\begin{enumerate}
    \item Empirical finding. We identify that prior citation predictors overfit explicit signals, leading to degradation on long-tail papers and under distribution shifts.  
    \item Multi-source Informed Graph Learning Framework. We propose a collaborative framework that combines agent-based fine-grained implicit feature extraction with graph representation learning, enabling robust prediction even without early citation information.  
    \item Bias-Aware GNN Learning. We design a robust mechanism that integrates Stage-A/Stage-B modeling, GroupDRO, and a regularization module, thereby suppressing superficial correlations, highlighting true correlations, and enhancing both accuracy and robustness.  
\end{enumerate}

\section{Related Work}

Research on scientific impact prediction and citation-network modeling falls into three directions: early-cascade modeling, metadata-driven graph learning, and dynamic or contrastive graph approaches.

\textbf{Early-cascade models} frame citation accumulation as a diffusion process. DeepCas~\citep{li2017deepcas} captures early propagation via random-walk citation paths and BiGRU attention, while SI-HDGNN~\citep{xu2022heterogeneous} embeds such cascades into heterogeneous academic graphs. Despite their effectiveness with sufficient citations, these models perform poorly in cold-start settings where early prediction is crucial.

\textbf{Metadata-driven graph models} leverage structural and textual metadata such as authors, venues, and topics. HINTS~\citep{10.1145/3442381.3450107} encodes dynamic heterogeneous networks with R-GCN and GRU, while CATE-HGNN~\citep{yang2023revisiting} and HLM-Cite~\citep{hao2024hlm} enrich semantics via clustering and pretrained language models. However, they remain sensitive to long-tail imbalance and often overfit superficial correlations.

\begin{figure*}[h]
  \centering
  \includegraphics[width=0.9\linewidth]{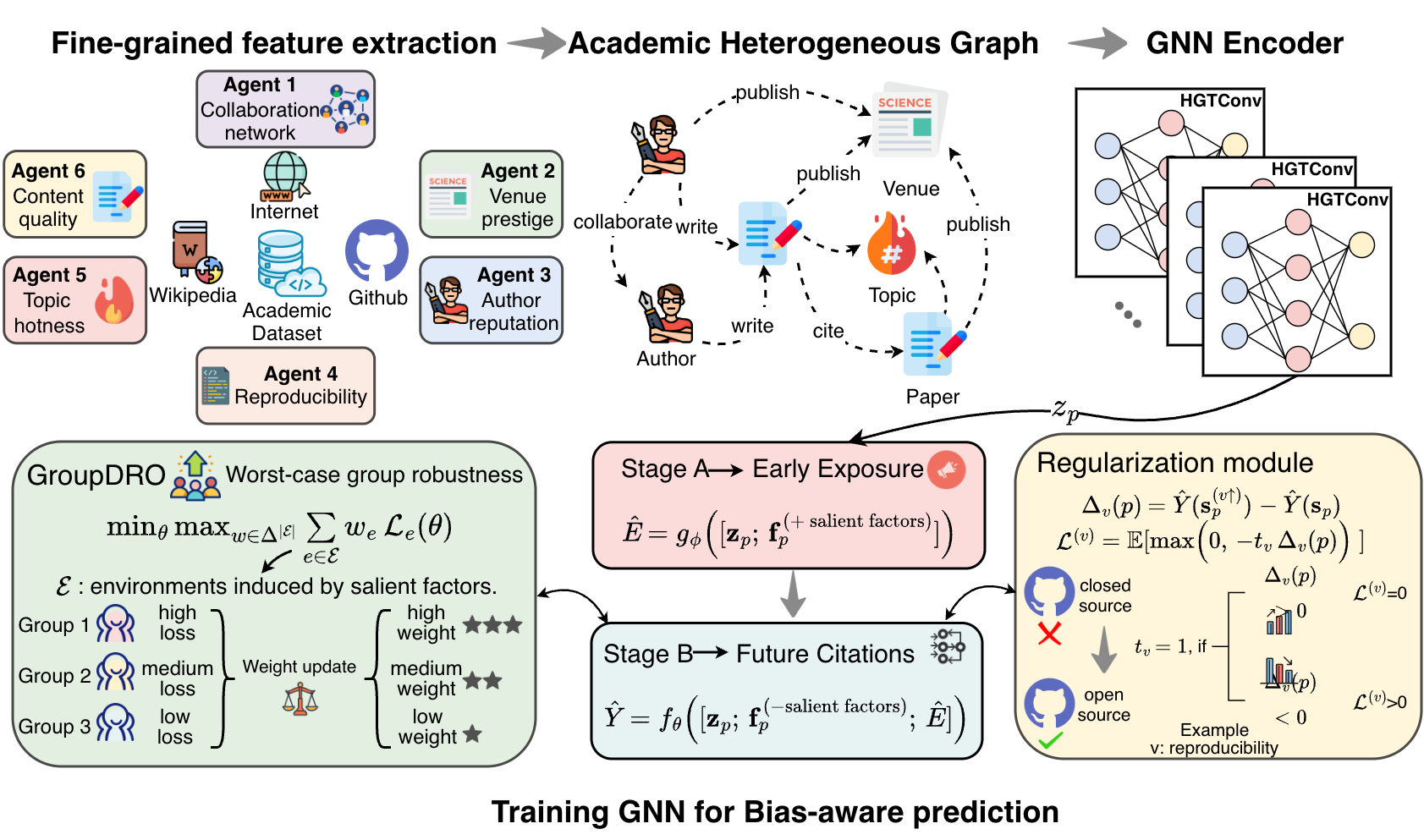}
  \caption{Overall architecture of the proposed \textit{BA-Cite} framework.}
  \label{fig:framework}
  \Description{}
\end{figure*}

\textbf{Dynamic and contrastive frameworks} model evolving citation contexts more explicitly.
H2CGL~\citep{he2023h2cgl} builds hierarchical heterogeneous graphs with citation-aware GIN, relation-aware GAT, and contrastive learning to integrate structural and temporal cues.
Related work such as TGN-TRec~\citep{shen2024temporal}, NETEVOLVE~\citep{miyake2024netevolve}, and ResearchTown~\citep{yu2024researchtown} explores dynamic or agent-based paradigms emphasizing interpretability and network evolution.
More recently, From Words to Worth~\citep{zhao2025words} proposes newborn article impact prediction, showing that fine-tuned LLMs can infer normalized impact (TNCSISP) from titles and abstracts alone, achieving competitive performance without citation history or external metadata.

\textbf{LLM-based semantic feature extraction.}
Recent studies leverage large language models (LLMs) to extract high-level semantic features\citep{wu2025leveraging,deng2026ai,wu2025scaling,li2025survey,yangadapting} from titles, abstracts, and metadata for scholarly impact prediction. LLMs provide contextualized representations and latent signals such as novelty or interdisciplinarity, which are incorporated into downstream graph models as auxiliary embeddings\citep{CHEN2026104336,jia-etal-2025-hetgcot,yangTdistributedSphericalFeature2023}, especially under cold-start settings. However, these features are typically treated as static enhancements, without explicitly modeling their interaction with dynamic graph structures or citation bias.

Despite these advances, few studies examine the mechanisms and bias dynamics driving citation disparities. We address this gap with a unified multi-source graph learning framework, enabling robust citation prediction under cold-start and distribution shifts.

\vspace{-2pt}
\section{Methodology}
In this section, we present our framework, \textit{BA-Cite}, illustrated in Fig.~\ref{fig:framework}. It consists of two parts: (i) agent-based fine-grained feature extraction, where multiple agents derive implicit factors from metadata and external resources; and (ii) graph learning on dynamic heterogeneous networks, where a GNN encoder integrates these signals through a two-stage forward process with bias- and robustness-oriented objectives.
In the following parts, we first outline the motivation, then describe the functions of individual agents, and finally detail the three GNN modules that incorporate agent-derived features into bias-aware representation learning.

\subsection{Motivation}
\subsubsection{Addressing Bias in Long-Tailed Citation Prediction}
In citation prediction, the inherently long-tailed distribution creates a persistent imbalance: a few highly cited papers dominate the learning process, while the majority receive limited attention.
This skew causes models to produce inaccurate predictions—often overestimating lowly cited papers and failing to generalize under distribution shifts, such as when predicting citations for evaluating new venues or identifying emerging research topics.
Such instability weakens the predictive reliability and limits the model’s ability to capture the real scientific impact across domains and time, for example, leading to biased predictive results that underestimate the contributions of early-career researchers and non-mainstream research. 
Addressing bias under the long-tailed and shifting distributions is therefore essential not only for improving prediction robustness but also for promoting fairness and inclusiveness in scientific assessment.

\subsubsection{Capturing Core Factors Driving Citation Dynamics with Agents and GNN(s)}
Existing citation prediction models often rely on handcrafted metadata or early citation signals, which are unavailable for newborn papers in cold-start settings and difficult to model manually.
Agents, by contrast, can autonomously mine and reason over multi-source information — such as reproducibility, topic hotness, and collaboration network — to derive fine-grained, high-level semantic factors that are otherwise implicit or sparsely encoded in metadata.
However, semantic cues alone cannot capture the structural and temporal dependencies that shape citation dynamics. While agents excel at extracting implicit knowledge, GNNs effectively model heterogeneous academic networks. Combining them enables semantic knowledge to be structurally grounded, ensuring both predictive accuracy and robustness.

\subsection{Fine-grained feature extraction}
\label{sec:feature}
To enrich paper representations, we design six domain-specific agents that automatically derive implicit features from metadata and external resources. Each agent focuses on a distinct dimension of citation dynamics, leveraging domain heuristics and multi-source knowledge to uncover fine-grained semantic cues that are difficult to encode manually. Together, these agents capture complementary aspects such as author reputation, venue prestige, collaboration patterns, reproducibility, topic hotness, and text quality, thereby expanding the feature space and enabling more balanced and generalizable representations. Given a paper $p$, the outputs are concatenated into a unified feature vector $\mathbf{f}_p = [A, V, R, C, H, Q]$, where each component corresponds to one agent's extracted factor. Below we describe the reason for choosing these factors and how each agent extracts them from metadata and external sources.

\paragraph{\textbf{Role-aware Author Reputation (A)}} Readers tend to cite works by reputable scholars or rising researchers, yet an author’s position within the byline also matters—first authors indicate primary contribution, last authors reflect senior leadership, while middle authors exert weaker influence~\citep{petersen2014reputation}.

\noindent\textit{Extraction process.} We assess author reputation by partitioning the author list into three roles: first author, last author, and other co-authors. For each role, we retrieve metadata such as institutional affiliation, publication count, and total citations; institutional prestige is further enriched via external sources (e.g., Wikipedia). These signals are aggregated into a continuous score on a 1–5 scale. During training, the scores of different roles are assigned different weights to reflect their varying influence on citation outcomes.

\paragraph{\textbf{Venue Prestige (V)}} Prestigious venues act as credibility signals, making their publications more visible and trusted, hence more likely to be cited~\citep{callaham2002journal}.

\noindent\textit{Extraction process.} We assess venue prestige by matching the venue name against external rankings (China Computer Federation Recommended Rankings (CCF) and Computing Research and Education Association of Australasia (CORE)) using both exact and fuzzy matching. The agent outputs a score on a 1–5 scale.

\paragraph{\textbf{Reproducibility (R)}} Open-source code or data enhances transparency, and reuse, driving credibility and long-term impact~\citep{raff2023does}.

\noindent\textit{Extraction process.} We assess reproducibility by scanning the content for open-source indicators (e.g., GitHub/GitLab links). If links are detected, we verify whether the repository contains code or datasets. The agent outputs a binary score (0/1).

\paragraph{\textbf{Collaboration Network (C)}} Broad and diverse collaborations, especially across institutions or countries, increase attention and citation potential through higher credibility and dissemination~\citep{wu2019large}.

\noindent\textit{Extraction process.} We assess collaboration characteristics, including team size, institutional diversity, and cross-country collaboration. Institutional metadata are complemented with external lookups (e.g., Wikipedia) to estimate prestige and geographic dispersion. The agent outputs a composite score on a 1–5 scale.

\paragraph{\textbf{Topic Hotness (H)}} Work in trending or growing areas gains citations faster by aligning with the community’s interests~\citep{wei2013scientists}.

\noindent\textit{Extraction process.} We assess topical hotness using the paper’s keywords. For each keyword, we count the number of papers in the previous year; the mean count across keywords is used as the hotness score. The agent outputs a continuous value.

\paragraph{\textbf{Text Quality (Q)}} Clear, well-structured titles and abstracts improve readability, directly influencing citation outcomes~\citep{jin2021research}.

\noindent\textit{Extraction process.} We assess text quality by prompting an LLM with the paper’s title and abstract, together with best-paper exemplars as references. The LLM evaluates structural clarity and professional expression and produces a score on a 1–5 scale.

\paragraph{Output}
The six features are concatenated into $\mathbf{f}_p$ and injected as attributes of the paper node $p$ in the heterogeneous academic graph. These enriched node features are then propagated through the GNN encoder, enabling the model to jointly capture structural patterns and implicit semantic signals.

\subsection{Bias-Aware GNN Learning Modules}

\begin{figure}[t]
  \centering
  \includegraphics[width=\linewidth]{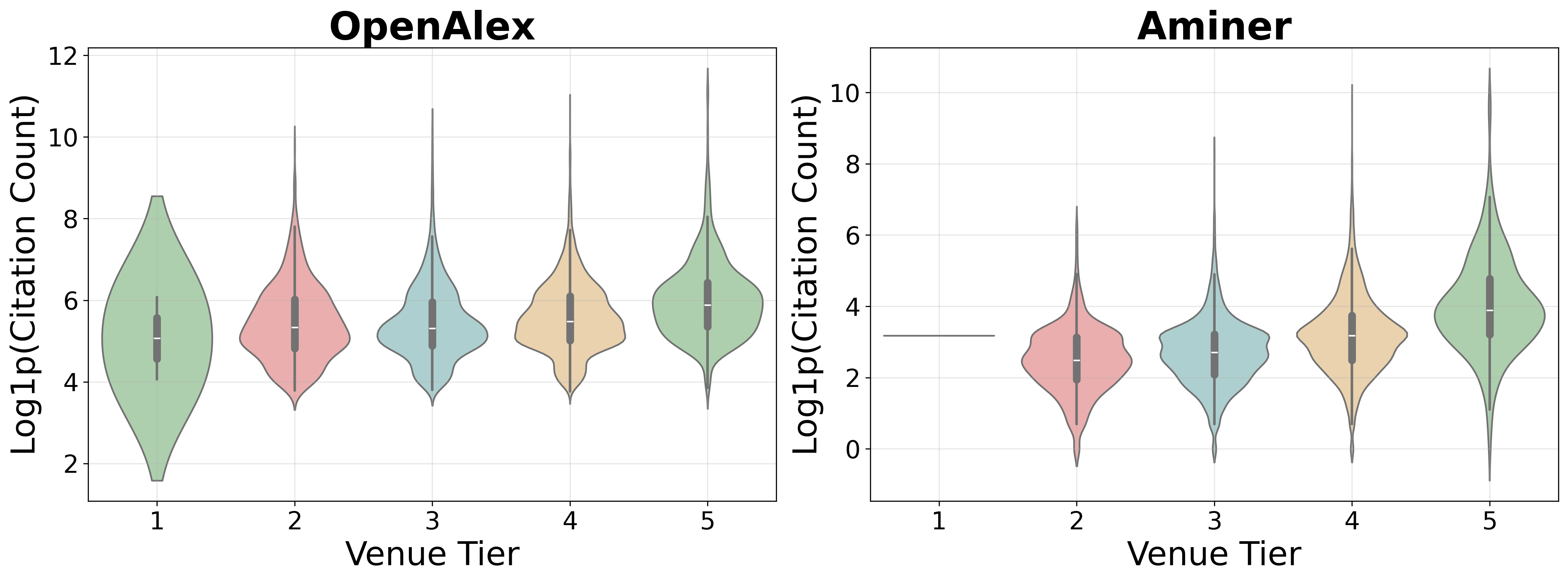}
   \caption{Distribution of citation counts across venues of different prestige levels on the Aminer and OpenAlex datasets.}
  \label{fig:venue}
  \Description{}
\vskip -0.1in
\end{figure}

With agent-derived features injected as paper-node attributes in the heterogeneous graph, the GNN encoder propagates them together with structural and relational information to form enriched paper representations. A key challenge is that conventional metadata-based models often overfit explicit correlations such as venue, which reflect distributional skew rather than true importance. This shortcut undermines prediction on high-impact papers outside top venues and widens gaps on lowly cited cases. To address these issues, we introduce three complementary modules: 
(i) a two-stage predictor that first estimates an intermediate exposure variable from graph embeddings and agent-derived features, and then predicts future citations based on both the exposure estimate and remaining signals;  
(ii) a GroupDRO objective applied to the prediction loss, which minimizes worst-group risk and alleviates head-dominated bias; and  
(iii) a counterfactual intervention head that perturbs controllable factors and regularizes predictions through monotonicity and smoothness constraints.
Together, these modules reduce over-reliance on explicit venue cues, leverage implicit drivers of scientific impact, and yield more robust citation predictions.

\subsubsection{Two-Stage Forward Strategy for Fair Representation}
Venue serves as the most salient shortcut feature in citation prediction: it is strongly correlated with citation counts but does not constitute a true determinant~\citep{uzzi2013atypical,10.1145/3442381.3450107,fortunato2018science}. As shown in Fig.~\ref{fig:venue}, higher-ranked venues typically exhibit higher average citation counts, but individual papers still show wide variation in citations. In computer science, this distributional pattern often arises because top venues cluster popular topics, attract well-known researchers, and also a larger proportion of papers release open-source code. A robust predictor should therefore capture the joint influence of multiple factors on future citations rather than being misled by a single explicit signal. To this end, we isolate venue effects by enforcing the influence pathway $V \!\to\! E \!\to\! Y$, where $E$ denotes \emph{early exposure} and $Y$ denotes future citations. A heterogeneous graph encoder first produces paper embeddings; Stage~A estimates $E$ using venue (among other features), while Stage~B predicts $Y$ without direct venue input, letting $V$ influence $Y$ only through $\hat E$.

\paragraph{Stage A: Exposure Estimation (\(V \rightarrow E\)).}
Stage~A estimates the latent early exposure variable \(\hat{E}\) from both the 
graph encoder and metadata features. The encoder operates on the full heterogeneous 
graph that includes venue nodes, so the paper embedding \(\mathbf{z}_p\) 
already incorporates venue effects. Together with the venue-inclusive feature vector 
\(\mathbf{f}_p^{(+V)} = [V, R, C, H, Q, Y, A_1, A_2, A_3]\), 
which includes venue prestige, reproducibility (\(R\)), collaboration network (\(C\)), 
topic hotness (\(H\)), text quality (\(Q\)), publication year (\(Y\)), 
and differentiated author reputations for first (\(A_1\)), last (\(A_2\)), 
and other co-authors (\(A_3\)), these signals are passed to a feed-forward head 
\(g_{\phi}\) to produce \(\hat{E}\).

\paragraph{Stage B: Venue-Excluded Prediction (\(E \rightarrow Y\)).}
Stage~B predicts the final citation count using a simplified graph where venue 
nodes and edges are removed, yielding a venue-excluded paper embedding. The input 
to the predictor \(f_{\theta}\) is the concatenation of this embedding, the 
venue-excluded feature vector \(\mathbf{f}_p^{(-V)}\), and the Stage~A estimate 
\(\hat{E}\). In this way, venue influences the outcome only indirectly via 
exposure, enforcing the influence path \(V \rightarrow E \rightarrow Y\).

\paragraph{Log-MSE Output and Prediction Loss.}
Instead of a negative-binomial likelihood, we adopt a mean squared error (MSE) 
objective after applying a logarithmic transformation to citation counts in order 
to mitigate the heavy-tailed distribution. Specifically, the predictor outputs 
\(\hat{Y} = f_{\theta}\!\left([\mathbf{z}_p;\,\mathbf{f}_p^{(-V)};\,\hat{E}]\right)\), 
and the loss is defined as
\begin{equation}
\mathcal{L}_{\mathrm{pred}} \,=\, 
\left(\log(1+y_p) - \log(1+\hat{Y})\right)^2,
\label{eq:pred_loss}
\end{equation}
where \(y_p\) denotes the observed citation count of paper \(p\).
\paragraph{Module Discussion} 
In this module, we restructure the prediction pipeline into two stages:
\[
\text{Stage A: } 
\hat{E} = g_\phi([\mathbf{z}_p;\, \mathbf{f}_p^{(+V)}]), 
\quad
\text{Stage B: } 
\hat{Y} = f_\theta([\mathbf{z}_p;\, \mathbf{f}_p^{(-V)};\, \hat{E}]),
\]
Stage~B does not directly observe $V$, 
so the only pathway is $V \to E \to Y$.  
From an information-theoretic view,
$ \
I(S;Y\mid \hat{E}) < I(S;Y),
\ $
meaning the shortcut influence of $S$ on $Y$ is strictly reduced and the model is forced to rely more on other implicit drivers.

\subsubsection{Environment-Aware Optimization for Robust Generalization}
We apply GroupDRO~\citep{sagawa2019distributionally} to the Stage-B prediction loss to prioritize the worst environment and improve cross-environment generalization. 
To mitigate the venue-dominated shortcut, we partition the training data into two environments by venue tier, $\mathcal{E}=\{\text{low},\,\text{high}\}$. 
For environment $e\in\mathcal{E}$ with index set $\mathcal{D}_e$, 
the group risk is defined as the mean loss on samples from that environment:
\begin{equation}
\mathcal{L}_e(\theta)
= \frac{1}{|\mathcal{D}_e|} \sum_{i \in \mathcal{D}_e} 
\ell\big(f_\theta(x_i), y_i\big),
\end{equation}
where $f_\theta(x_i)$ denotes the model prediction for paper $i$, $y_i$ is the observed citation count, and $\ell(\cdot)$ is the \emph{Stage-B} prediction loss defined in Eq.~\ref{eq:pred_loss} (Log-MSE on $\log(1+y)$).

\paragraph{GroupDRO objective.}
We optimize a worst-group risk via adversarial reweighting:
\begin{equation}
\min_{\theta}\;\max_{w\in\Delta^{2}}\;
\sum_{e\in\{\text{low},\,\text{high}\}} w_e\,\mathcal{L}_e(\theta),
\quad
\Delta^{2}=\left\{
\begin{array}{l}
w_{\text{low}},w_{\text{high}}\ge0, \\
w_{\text{low}}+w_{\text{high}}=1
\end{array}
\right\},
\end{equation}
where $w=(w_{\text{low}},w_{\text{high}})$ denotes nonnegative environment weights lying in the probability simplex $\Delta^2$.  
The inner maximization allocates more weight to the environment with larger risk, forcing the model to improve performance on the worst-performing group.

\paragraph{Weight update.}
Let $\bar{\mathcal{L}}=\tfrac{1}{2}\sum_e \mathcal{L}_e$ and 
$\sigma_{\mathcal{L}}=\sqrt{\tfrac{1}{2}\sum_e (\mathcal{L}_e-\bar{\mathcal{L}})^2}$ 
(the population standard deviation of group risks). Here $\mathcal{L}_e$ is the average loss of environment $e$, $\bar{\mathcal{L}}$ is the mean loss across environments, and $\sigma_{\mathcal{L}}$ is their standard deviation. With step size $\alpha>0$ and small constant $\varepsilon$,
\begin{equation}
\tilde{w}_e \;\propto\; 
w_e^{(t)} \exp\!\Big(\alpha\,\frac{\mathcal{L}_e-\bar{\mathcal{L}}}{\sigma_{\mathcal{L}}+\varepsilon}\Big),
\qquad 
\sum_e \tilde{w}_e = 1,
\end{equation}
where $w_e^{(t)}$ is the current weight of environment $e$. Higher-than-average loss increases the weight, while lower loss decreases it. Finally, weights are clamped and renormalized:
\begin{equation}
\hat{w}_e=\mathrm{clip}(\tilde{w}_e,\,w_{\min},\,w_{\max}),
\qquad
w_e^{(t+1)}=\frac{\hat{w}_e}{\sum_{e'}\hat{w}_{e'}}.
\end{equation}

Here $[w_{\min},w_{\max}]$ bounds prevent degenerate values, and $w_e^{(t+1)}$ denotes the updated weight for environment $e$.

\paragraph{Module Discussion}
This environment-aware optimization prevents the model from collapsing onto dominant groups shaped by highly cited or high-prestige papers. 
By forcing improvements on underrepresented environments, GroupDRO enhances fairness, promoting robust generalization across diverse citation contexts.

\subsubsection{Regularization for Reasonable and Stable Prediction.}
To provide actionable “what-if’’ estimates, we augment Stage~B with a regularization head. 
This module turns abstract features into interpretable regularization effects: it quantifies the predicted citation change if \emph{only} a controllable factor $v$ were improved (e.g., toggling $R\!:\!0\!\to\!1$), while all other attributes remain fixed and the induced change in early exposure is propagated consistently. 
This yields predictions that are both actionable and constrained to be directionally reasonable and stable.

For a controllable factor $v$ (e.g., reproducibility $R$), let $\mathbf{s}_p=[\mathbf{z}_p;\,\mathbf{f}_p^{(-V)};\,\hat E]$ be the Stage~B input constructed from the observed features, and let $\mathbf{s}_p^{(v\uparrow)}$ be the same input after setting $v$ to a high value (keeping all other features fixed) and recomputing $\hat E$ under this change. 
The per-factor counterfactual effect is defined as
\begin{equation}
\Delta_v(p)\;=\; \hat Y\!\left(\mathbf{s}_p^{(v\uparrow)}\right)\;-\;\hat Y\!\left(\mathbf{s}_p\right),
\end{equation}
where $\hat Y(\cdot)=f_\theta(\cdot)$ and the early-exposure estimate in $\mathbf{s}_p^{(v\uparrow)}$ is $\hat E^{(v\uparrow)}=g_\phi([\mathbf{z}_p;\,\mathbf{f}_p^{(+V)}\!\text{ with }v\!\leftarrow\!\text{high}])$.

\paragraph{Monotonicity and smoothness regularization.}
Let $t_v\!\in\!\{+1,-1\}$ denote the expected direction of improvement (typically $t_v=+1$), 
and let $\tau_v$ be a threshold that marks the “low’’ region of $v$ (e.g., $R{=}0$). 
We enforce that raising $v$ should not hurt citations for low-value cases, 
and keep effects calibrated via a smoothness penalty:
\begin{equation}
\mathcal{L}_{\text{mono}}^{(v)} \;=\; 
\mathbb{E}\!\left[\,\max\!\big(0,\,-\,t_v\,\Delta_v(p)\big)\;\mathbf{1}\{v(p)<\tau_v\}\,\right],
\end{equation}
\begin{equation}
\mathcal{L}_{\text{smooth}}^{(v)} \;=\; 
\mathbb{E}\!\left[\,\Delta_v(p)^2\,\right].
\end{equation}

Aggregating over controllable factors $\mathcal{V}$, the total regularizer is
\begin{equation}
\mathcal{L}_{\text{reg}}
= \lambda_{\text{mono}} \sum_{v\in\mathcal{V}} \mathcal{L}_{\text{mono}}^{(v)}
\;+\;
\lambda_{\text{smooth}} \sum_{v\in\mathcal{V}} \mathcal{L}_{\text{smooth}}^{(v)}.
\end{equation}

\paragraph{Module Discussion}
One major source of bias in citation prediction arises from the limited and coarse metadata used in traditional models, which makes explicit factors dominate the learning process. 
Within our regularization module, the model leverages fine-grained, implicitly derived features to regularize representation learning, thereby mitigating shortcut reliance. 
By distributing explanatory power across multiple implicit drivers—such as collaboration patterns, topic dynamics, and text quality—the model becomes less dependent on any single explicit factor and achieves more balanced generalization across varying citation environments.

\subsubsection{Objective.}
Our training objective focuses on environment-aware risk (GroupDRO) and the proposed sensitivity regularization:
\begin{equation}
\mathcal{L}_{\text{total}}
= \lambda_{\text{main}}\,\mathcal{L}_{\text{groupdro}} 
+  \lambda_{\text{reg}}\,\mathcal{L}_{\text{reg}}.
\end{equation}

We also employ two lightweight auxiliaries on Stage~B—an exposure calibration loss on $\hat E$ and an adversarial venue-invariance loss—which are reported in ablations and described in Appendix~\ref{app:aux}, but omitted here for brevity.

\section{Experiments}
In this section, we conduct extensive experiments to answer the following four research questions:

\textbf{RQ1}: How does BA-Cite perform compared with other models in terms of predictive accuracy, and ranking quality?

\textbf{RQ2}: How does each component contribute to the overall performance of BA-Cite?

\textbf{RQ3}: Does BA-Cite achieve robust prediction across different data distributions after bias mitigation?

\textbf{RQ4}: How do different hyper-parameters affect BA-Cite ?

\subsection{Experimental Setup}
\subsubsection{Datasets.}
We select two widely used publicly available academic datasets, \textbf{Aminer}~\citep{tang2008arnetminer} and \textbf{OpenAlex}~\citep{priem2022openalex}, to verify the effectiveness of our proposed framework. 
Both datasets focus on the \textit{computer science} domain and contain heterogeneous nodes, as well as temporal relations such as \textit{cites}, \textit{writes}, and \textit{has\_topic}. 
The temporal coverage ranges from \textbf{2010 to 2025}. 

For each paper, the prediction target is its future citation number within the next five years (starting from the second year) after publication. 
We split the data by year, using papers published during 2010--2018 for training, 2019 for validation, and 2020 for testing. 
Within each split, we randomly sample 10,000 papers for training, 1,000 for validation, and 1,000 for testing. 
To reduce randomness, we repeat the sampling process three times and conduct five runs with different random seeds for each sample. 
The reported results include the mean and standard deviation across all runs.

\subsubsection{Baselines.}
We compare our framework with representative methods from four categories, covering classical GNNs, sequential models, large language models, and metadata-based neural models.

\noindent\textbf{Graph Neural Network-based methods.}
\begin{itemize}[leftmargin=12pt]
    \item \textbf{GAT (ICLR'18)}~\citep{velivckovic2017graph}: models citation relations using multi-head graph attention.
    \item \textbf{HINTS (WWW'21)}~\citep{10.1145/3442381.3450107}: encodes temporal heterogeneous information networks for citation time-series prediction.
    \item \textbf{DyGFormer (NeurIPS'23)}~\citep{yu2023towards}: applies transformer-style temporal encoding for dynamic graphs.
\end{itemize}

\noindent\textbf{Sequence-based methods.}
\begin{itemize}[leftmargin=12pt]
    \item \textbf{BiLSTM-Meta (Scientometrics'21)}~\citep{ma2021deep}: captures citation sequences via bidirectional recurrent modeling.
    \item \textbf{DeepCas (WWW'17)}~\citep{li2017deepcas}: learns citation cascade representations through random walks and BiGRU-based attention.
    \item \textbf{SI-HDGNN (KBS'22)}~\citep{xu2022heterogeneous}: builds heterogeneous dynamic academic networks for impact propagation.
\end{itemize}

\noindent\textbf{Large Language Model-based methods.}
\begin{itemize}[leftmargin=12pt]
    \item \textbf{GPT-4o (OpenAI'24)}~\citep{hurst2024gpt}: leverages LLM reasoning and knowledge for citation impact estimation.
    \item \textbf{Llama-3.1-405B (Meta'24)}~\citep{dubey2024llama}: employs open-source LLM embeddings for academic impact inference.
    \item \textbf{NAIP (AAAI'25)}~\citep{zhao2025words}: formulates newborn article impact prediction by fine-tuning large language models on title–abstract pairs with the TNCSISP metric, enabling content-only impact estimation without external metadata.
\end{itemize}

\noindent\textbf{Metadata-based Neural Methods.}
\begin{itemize}[leftmargin=12pt]
\item \textbf{BP-NN (J. Informetrics’20)}~\citep{ruan2020predicting}: 
A four-layer feed-forward neural network that predicts five-year citation counts. 

\end{itemize}

\noindent These baselines represent diverse paradigms in scientific impact prediction, from early cascade modeling to metadata-driven, dynamic, and LLM-enhanced frameworks, providing a comprehensive comparison foundation. 
For all feature-dependent baselines, we supply the fine-grained semantic features extracted in Section~\ref{sec:feature} to ensure consistent and enriched input representations.

\begin{table*}[t]
\centering
\caption{Comparison on Aminer and OpenAlex: MALE, RMSLE, NDCG@10, and NDCG@20 (mean $\pm$ std). Lower is better for MALE/RMSLE; higher is better for NDCG. Best results are in \textbf{bold}, second best are \underline{underlined}. “†” indicates statistically significant improvement over all baselines under a paired t-test with \( p < 0.05 \).}
\setlength{\tabcolsep}{4.2pt}
\begin{tabular}{lcccccccc}
\toprule
& \multicolumn{4}{c}{Aminer} & \multicolumn{4}{c}{OpenAlex} \\
\cmidrule(lr){2-5}\cmidrule(lr){6-9}
Baselines & MALE $\downarrow$ & RMSLE $\downarrow$ & NDCG@10 $\uparrow$ & NDCG@20 $\uparrow$ & MALE $\downarrow$ & RMSLE $\downarrow$ & NDCG@10 $\uparrow$ & NDCG@20 $\uparrow$ \\
\midrule
\multicolumn{9}{c}{\hspace{3em}\textit{Graph Neural Network-based methods}} \\
GAT       & 0.94 ± 0.02 & 1.09 ± 0.02 & 0.09 ± 0.11 & 0.10 ± 0.11 & \underline{0.87 ± 0.04} & 1.14 ± 0.03 & 0.23 ± 0.17 & 0.25 ± 0.16 \\
HINTS              & 0.99 ± 0.01 & 1.14 ± 0.01 & 0.03 ± 0.01 & 0.04 ± 0.01 & 0.98 ± 0.00 & 1.14 ± 0.00 & 0.04 ± 0.02 & 0.05 ± 0.03 \\
DyGFormer          & 1.36 ± 0.36 & 1.55 ± 0.38 & 0.27 ± 0.09 & 0.26 ± 0.07 & 1.09 ± 0.23 & 1.36 ± 0.24 & \underline{0.39 ± 0.06} & \underline{0.43 ± 0.05} \\
\midrule
\multicolumn{9}{c}{\hspace{3em}\textit{Sequence-based methods}} \\
BiLSTM        & 0.79 ± 0.02 & 0.98 ± 0.03 & \underline{0.32 ± 0.08} & \underline{0.33 ± 0.06} & 2.21 ± 0.11 & 2.35 ± 0.11 & 0.17 ± 0.06 & 0.13 ± 0.08 \\
DeepCas            & 0.90 ± 0.02 & 1.30 ± 0.03 & 0.04 ± 0.02 & 0.05 ± 0.02 & 1.09 ± 0.01 & 1.33 ± 0.01 & 0.06 ± 0.02 & 0.08 ± 0.03 \\
SI-HDGNN           & \underline{0.73 ± 0.02} & 1.01 ± 0.02 & 0.16 ± 0.10 & 0.17 ± 0.09 & 1.05 ± 0.04 & 1.31 ± 0.04 & 0.11 ± 0.07 & 0.05 ± 0.05 \\
\midrule
\multicolumn{9}{c}{\hspace{3em}\textit{Large Language Model-based methods}} \\
GPT-4o             & 1.73 ± 0.03 & 1.91 ± 0.04 & 0.13 ± 0.02 & 0.23 ± 0.07 & 0.90 ± 0.01 & \underline{1.12 ± 0.00} & 0.38 ± 0.00 & 0.36 ± 0.04 \\
Llama\_3.1\_405b   & 1.62 ± 0.01 & 1.80 ± 0.01 & 0.12 ± 0.02 & 0.18 ± 0.01 & 1.02 ± 0.08 & 1.23 ± 0.10 & 0.31 ± 0.09 & 0.34 ± 0.03 \\
NAIP\_Llama        & 0.77 ± 0.02 & \underline{0.97 ± 0.03} & 0.05 ± 0.01 & 0.06 ± 0.01 & 2.53 ± 0.01 & 2.73 ± 0.01 & 0.01 ± 0.01 & 0.01 ± 0.02 \\
NAIP\_Qwen         & 0.83 ± 0.02 & 1.04 ± 0.03 & 0.05 ± 0.02 & 0.06 ± 0.01 & 2.46 ± 0.01 & 2.66 ± 0.01 & 0.08 ± 0.03 & 0.10 ± 0.04 \\
\midrule
\multicolumn{9}{c}{\hspace{3em}\textit{Metadata-based Neural methods}} \\
BP-NN         & 0.84 ± 0.05 & 1.05 ± 0.06 & 0.27 ± 0.12 & 0.31 ± 0.12 & 2.25 ± 0.25 & 2.38 ± 0.24 & 0.23 ± 0.07 & 0.20 ± 0.04 \\
\midrule
\multicolumn{9}{c}{\hspace{3em}\textit{Our Method}} \\
BA-Cite            & \textbf{0.71 ± 0.02}$^{\dagger}$ & \textbf{0.88 ± 0.03}$^{\dagger}$ & \textbf{0.34 ± 0.12}$^{\dagger}$ & \textbf{0.37 ± 0.13}$^{\dagger}$ & \textbf{0.67 ± 0.01}$^{\dagger}$ & \textbf{0.92 ± 0.04}$^{\dagger}$ & \textbf{0.51 ± 0.04}$^{\dagger}$ & \textbf{0.47 ± 0.01}$^{\dagger}$ \\
\bottomrule
\end{tabular}
\label{tab:aminer-openalex-core-metrics}
\end{table*}

\begin{table*}[t]
\centering
\caption{Ablation study of BA-Cite on Aminer and OpenAlex: MALE, RMSLE, NDCG@10, and NDCG@20 (mean $\pm$ std).}
\setlength{\tabcolsep}{4.5pt}
\begin{tabular}{lcccccccc}
\toprule
& \multicolumn{4}{c}{Aminer} & \multicolumn{4}{c}{OpenAlex} \\
\cmidrule(lr){2-5}\cmidrule(lr){6-9}
Ablation Type & MALE $\downarrow$ & RMSLE $\downarrow$ & NDCG@10 $\uparrow$ & NDCG@20 $\uparrow$ & MALE $\downarrow$ & RMSLE $\downarrow$ & NDCG@10 $\uparrow$ & NDCG@20 $\uparrow$ \\
\midrule
\textit{w/o Feature}   & 0.86 ± 0.02 & 1.00 ± 0.03 & 0.10 ± 0.07 & 0.11 ± 0.06 & 1.88 ± 0.00 & 2.04 ± 0.02 & 0.07 ± 0.00 & 0.09 ± 0.01 \\
\textit{w/o Two-Stage} & \textbf{0.70 ± 0.02} & \textbf{0.87 ± 0.02} & \underline{0.26 ± 0.18} & \underline{0.29 ± 0.16} & \underline{0.78 ± 0.03} & \underline{0.99 ± 0.05} & 0.16 ± 0.08 & 0.19 ± 0.07 \\
\textit{w/o Reg} & 0.78 ± 0.02 & 0.97 ± 0.03 & 0.14 ± 0.08 & 0.14 ± 0.08 & 1.21 ± 0.20 & 1.30 ± 0.18 & \underline{0.18 ± 0.09} & \underline{0.21 ± 0.07} \\
\textit{w/o GroupDRO}  & 0.79 ± 0.02 & 0.89 ± 0.03 & 0.01 ± 0.00 & 0.03 ± 0.01 & 0.99 ± 0.02 & 1.09 ± 0.03 & 0.05 ± 0.02 & 0.07 ± 0.03 \\
BA-Cite            & \underline{0.71 ± 0.02} & \underline{0.88 ± 0.03} & \textbf{0.34 ± 0.12} & \textbf{0.37 ± 0.13} & \textbf{0.67 ± 0.01} & \textbf{0.92 ± 0.04} & \textbf{0.51 ± 0.04} & \textbf{0.47 ± 0.01} \\
\bottomrule
\end{tabular}
\label{tab:ablation-core-metrics}
\end{table*}

\subsubsection{Implementation Details.}
We implement the counterfactual two-stage HGT in PyTorch. The heterogeneous encoder uses two HGT layers (hidden size $128$, $4$ attention heads, dropout $0.4$). Node features follow our schema: paper nodes have an $8$-dimensional vector $[R,Q,C,H,Y,A_1,A_2,A_3]$, while author/venue/topic nodes are initialized with $1$-dimensional metadata features.
For counterfactual learning, we enable \emph{reproducibility} $(R)$ and \emph{content quality} $(Q)$ as actionable variables and apply monotonicity regularization so that larger $(R,Q)$ should not decrease predicted citations. We also employ adversarial training and an auxiliary loss.
We adopt GroupDRO over $2$ environments with step size $\alpha=0.1$ and group-weight clipping to $[0.1,0.9]$. Environments are constructed by venue prestige threshold $\tau=0.8$.
Optimization uses AdamW (lr $=10^{-3}$, weight decay $=10^{-4}$) with a 10-epoch warm-up followed by cosine decay to $10^{-5}$. We train up to $200$ epochs with batch size $128$ and early stopping on validation loss.

\subsubsection{Evaluation Metrics.}
We evaluate model performance from two perspectives: 
(i) point accuracy using Mean Absolute Log Error (MALE) and Root Mean Squared Log Error (RMSLE); and 
(ii) ranking quality using Normalized Discounted Cumulative Gain (NDCG@K, $K{=}10,20$)~\citep{jarvelin2002cumulated}. 
All log-based metrics adopt $\log(1{+}y)$ to mitigate heavy-tailed citation counts. Lower is better for MALE/RMSLE, while higher is better for NDCG. The detailed calculation formulas are provided in the Appendix~\ref{sec:experiments}.

\begin{figure*}[t]
  \centering
  \includegraphics[width=\linewidth]{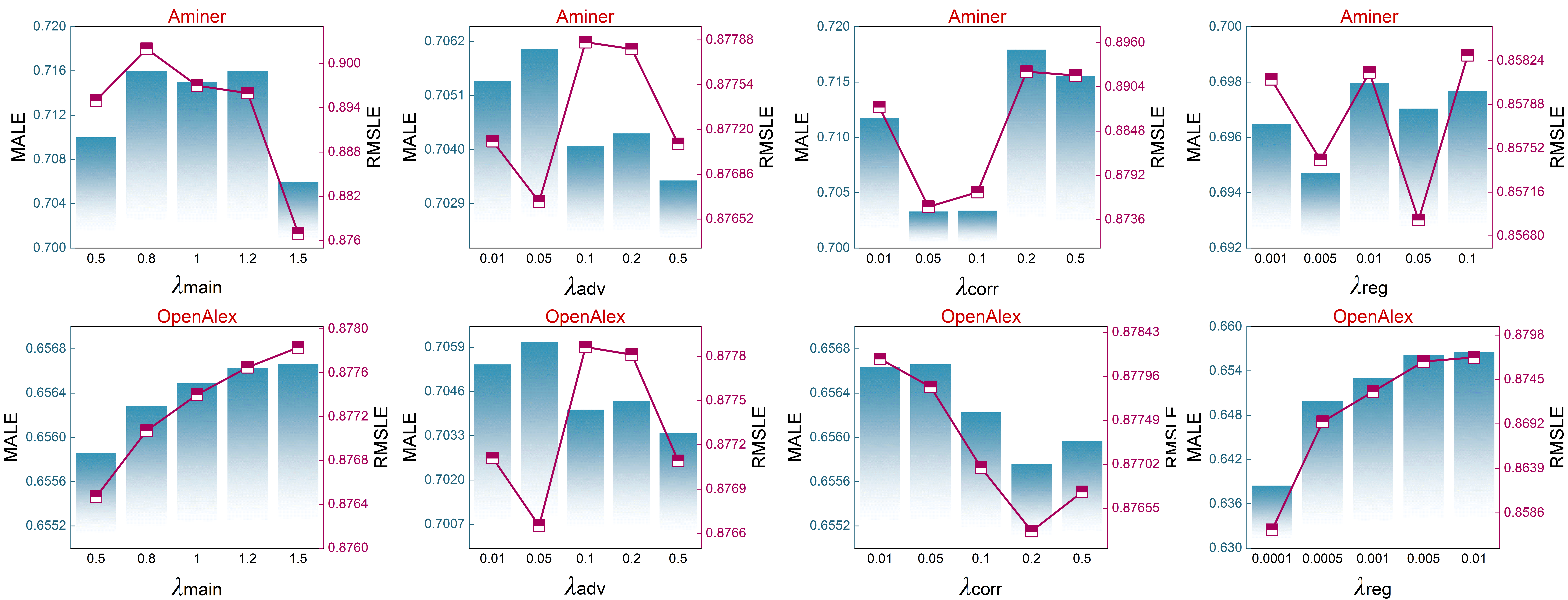}
  \caption{Parameter sensitivity results of \textit{BA-Cite} on the Aminer and OpenAlex datasets.}
  \label{fig:parameter}
  \Description{}
\end{figure*}

\begin{figure}[h]
  \centering
  \includegraphics[width=\linewidth]{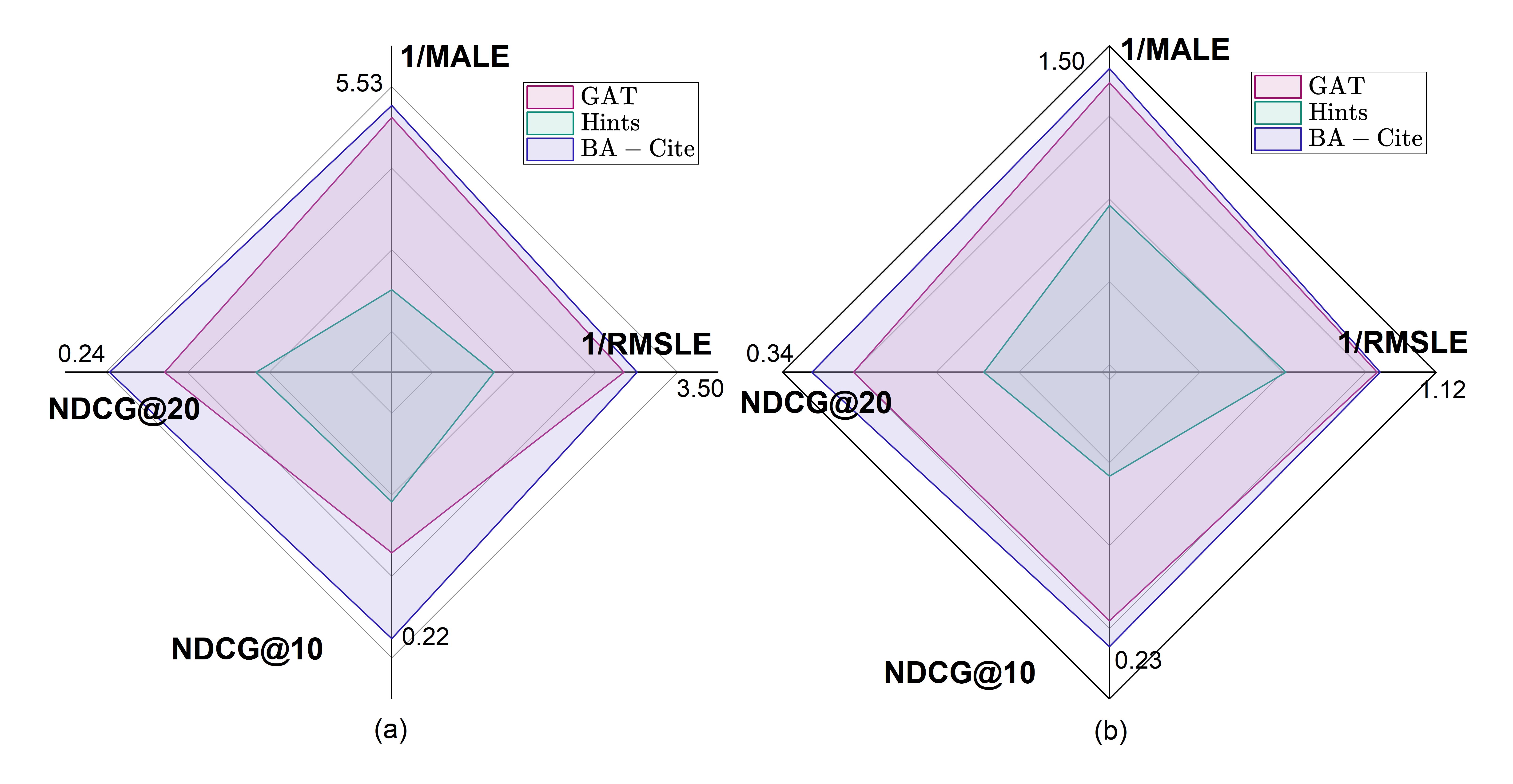}
  \caption{Performance comparison of three models on Aminer dataset with different citation levels. 
(a) Results on lowly cited papers. 
(b) Results on highly cited papers.}
  \label{fig:parameter1}
  \Description{}
\vskip -0.2in
\end{figure}

\subsection{Overall Performance (RQ1)}
Table~\ref{tab:aminer-openalex-core-metrics} summarizes the overall results on the Aminer and OpenAlex datasets. In general, BA-Cite achieves the best performance across all metrics. On Aminer, it outperforms the best baselines by reducing MALE by 2.7\% and RMSLE by 9.3\%, while improving NDCG@10 and NDCG@20 by 2\% and 4\%. On OpenAlex, BA-Cite further reduces MALE by 20.3\% and RMSLE by 20.1\%, and boosts NDCG@10 and NDCG@20 by 12\% and 4\%. These consistent gains demonstrate that integrating agent-derived fine-grained features with dynamic heterogeneous graph learning effectively improves both accuracy and ranking quality. Compared to other methods, GNN-based models generally perform better than sequence- or metadata-based baselines, highlighting the importance of structural and temporal modeling. LLM-based methods achieve competitive ranking performance, which may benefit from pretrained knowledge rather than structural understanding. Overall, BA-Cite provides balanced and stable improvements, confirming its advantage in handling cold-start and long-tailed citation prediction scenarios.

\subsection{Ablation Study (RQ2)}
We examine the effect of each module in BA-Cite on Aminer and OpenAlex (Table \ref{tab:ablation-core-metrics}).
Removing fine-grained feature extraction (\textit{w/o Feature}) leads to the largest drop, confirming the importance of agent-derived semantic cues.
Eliminating the two-stage process (\textit{w/o Two-Stage}) also reduces ranking quality, showing that staged optimization enhances representation robustness. However, on the relatively unbiased Aminer dataset (as evidenced in Fig.~\ref{fig:venue}), the two-stage mechanism introduces slight structural redundancy and optimization noise, which marginally reduces fitting efficiency despite its advantage in bias mitigation.
Without bias regularization (\textit{w/o Reg}), both MALE and RMSLE slightly increase, indicating reduced robustness.
The absence of environment-aware training (\textit{w/o GroupDRO}) causes severe degradation in NDCG, proving its necessity for balanced performance across domains.
Overall, BA-Cite achieves robust and balanced performance, demonstrating that both agent-based feature extraction and bias-aware two-stage optimization are crucial for accurate and fair citation prediction.

\subsection{Analysis of Robustness (RQ3)}
As shown in Fig.~\ref{fig:parameter1}, we further evaluate whether \textit{BA-Cite} maintains stable performance after bias mitigation under varying data distributions. 
Following the partitioning strategy in~\citep{10.1145/3442381.3450107}, we divide papers into lowly cited and highly cited subsets by citation counts. 
Across all three compared models, \textit{BA-Cite} achieves the best performance, showing strong results on both low- and high-citation papers, which indicates its robustness to distributional variation.

\subsection{Analysis of Parameter Sensitivity (RQ4)}
As shown in Fig.~\ref{fig:parameter}, we analyze the sensitivity of four loss weights.
(1) $\lambda_{main}$. On Aminer, errors decrease with larger $\lambda_{main}$, peaking at 1.2–1.5, indicating that a stronger main objective improves fit under mild bias. On OpenAlex, MALE and RMSLE increase with $\lambda_{main}$, so smaller values (0.5–0.8) are preferable to preserve capacity for debiasing.
(2) $\lambda_{adv}$. A small-to-moderate adversarial strength works best: Aminer achieves its lowest errors around 0.05, while OpenAlex shows a trade-off—MALE near 0.05 and RMSLE near 0.2—suggesting $\lambda_{adv}$ in 0.05–0.2. Larger values introduce instability without gains.
(3) $\lambda_{corr}$. Both datasets exhibit a U-shaped trend: mild regularization helps (optimum around 0.1–0.2), whereas overly weak or strong settings (e.g., 0.01 or 0.5) degrade performance by allowing redundancy or over-constraining embeddings.
(4) $\lambda_{reg}$. Fairness regularization should be conservative. Aminer reaches minimum errors around 0.05, while OpenAlex favors very small values. Strong regularization degrades accuracy on both datasets.

\section{Conclusion}
We present BA-Cite, a bias-aware citation prediction framework combining multi-agent semantic extraction with dynamic heterogeneous graph learning.
By modeling author, venue, and topic dynamics, BA-Cite provides a strong semantic basis for citation reasoning.
Its two-stage optimization with GroupDRO enhances robustness and mitigates overfitting to high-prestige environments.
Experiments on Aminer and OpenAlex show consistent improvements over strong baselines, with stability confirmed by ablation and sensitivity analyses.
BA-Cite generalizes well under distribution shifts, supporting real-world scholarly impact evaluation.
Future work includes impact explanation, cross-domain transfer, and reinforcement learning–based adaptive bias mitigation.


\bibliographystyle{ACM-Reference-Format}

\bibliography{main}


\appendix

\section{Experiments Details}
\label{sec:experiments}

\noindent\textbf{Point accuracy.} 
Given $n$ papers with ground-truth five-year citations $y_i$ and predictions $\hat{y}_i$, we compute:
\begin{equation}
\mathrm{MALE} = \frac{1}{n} \sum_{i=1}^{n} \big|\log(1+\hat{y}_i) - \log(1+y_i)\big|,
\end{equation}
\begin{equation}
\mathrm{RMSLE} = \sqrt{\frac{1}{n} \sum_{i=1}^{n} \big(\log(1+\hat{y}_i) - \log(1+y_i)\big)^2},
\end{equation}

\noindent\textbf{Ranking quality.} 
For papers published in the same year, let $\pi$ denote the predicted ranking and $\pi^*$ the ground-truth ranking:
\begin{equation}
\mathrm{NDCG@K} = 
\frac{\sum_{i=1}^{K} \frac{y_{\pi(i)}}{\log_2(i+1)}}{
\sum_{i=1}^{K} \frac{y_{\pi^*(i)}}{\log_2(i+1)}}.
\end{equation}

\section{Auxiliary Losses for Stage~B}
\label{app:aux}

To further stabilize optimization and mitigate residual bias, we incorporate two lightweight auxiliary objectives at Stage~B: \textbf{(i) exposure calibration loss} and \textbf{(ii) adversarial venue-invariance loss}.  
Both are designed to regularize the learned citation representations without introducing additional parameters or inference overhead.

\subsection{Exposure Calibration Loss}
Empirical studies have shown that citation counts are strongly correlated with \textit{exposure factors}—such as publication venue, collaboration size, or open-source visibility—which may distort predictive learning.  
To prevent the model from over-amplifying these factors, we impose an auxiliary calibration constraint on the predicted exposure score~$\hat{E}$:
\begin{equation}
\mathcal{L}_{\text{exp}} = \mathrm{KL}\big( p(\hat{E}) \, \Vert \, p(E^*) \big),
\end{equation}
where $E^*$ denotes the empirical exposure distribution estimated from the training data.  
This term penalizes deviations between the predicted and empirical exposure distributions, ensuring that the model’s intermediate exposure estimation remains statistically consistent and well-calibrated.

\subsection{Adversarial Venue-Invariance Loss}
Venue prestige is one of the most dominant shortcut features in citation prediction.  
To enhance robustness against venue bias, we introduce an adversarial objective that enforces venue-invariant latent representations.  
A discriminator $D_v$ is trained to predict the venue label from the Stage~B feature embedding $\mathbf{s}_p$, while the main encoder $f_\theta$ attempts to fool it:
\begin{equation}
\min_{\theta} \max_{\phi} \; \mathbb{E}_{p \sim \mathcal{D}}
\big[ \log D_{v_\phi}(v_p \mid f_\theta(\mathbf{s}_p)) \big].
\end{equation}
This min–max game discourages the encoder from encoding venue-specific artifacts, resulting in a fairer and more transferable citation representation.

\subsection{Implementation Note}
Both auxiliaries are assigned small weights ($\lambda_{\text{exp}} = 0.1$, $\lambda_{\text{adv}} = 0.05$) relative to the primary Stage~B objective.  
They are only active during training and disabled during inference.  
Ablation results in Table~\ref{tab:ablation-core-metrics} confirm that incorporating these auxiliaries improves fairness and stability, particularly under long-tailed or domain-shifted scenarios.

\section{Dataset Diagnostics and Descriptive Statistics}
\label{app:data-diagnostics}

This appendix reports basic diagnostics of the evaluation splits used in our experiments, including overall scale and central tendency of citation counts, temporal coverage, authorship statistics, and venue/document-type compositions. These summaries help contextualize the long-tailed nature of citations and the salience of venue as a shortcut factor discussed in the main text.

\subsection{Global Characteristics}
\label{app:data:global}
Table~\ref{tab:dataset_characteristics_app} summarizes dataset-level statistics. We report the number of papers, citation central tendencies (mean/median/max), year range, average number of authors, and the highest venue tier present in each split.


\subsection{Venue Tier Distribution}
\label{app:data:venue}
Table~\ref{tab:venue_distribution_app} reports the venue-tier composition for each split. Percentages are computed over all papers in the split.

\begin{table}[H]
\centering
\caption{Dataset characteristics across splits.}
\label{tab:dataset_characteristics_app}
\setlength{\tabcolsep}{2pt}
\begin{tabular}{lccccccc}
\toprule
\textbf{Dataset} & \textbf{Avg Cit.} & \textbf{Max Cit.} & \textbf{Avg} & \textbf{Top} \\
& & & \textbf{Authors} & \textbf{Venue Tier} \\
\midrule
Aminer1           & 49.4  & 18{,}816 & 3.5 & Tier 4 \\
Aminer2           & 47.9  & 9{,}517  & 3.5 & Tier 4 \\
Aminer3           & 47.7  & 33{,}132 & 3.6 & Tier 4 \\
OpenAlex1  & 442.2 & 71{,}274 & 4.3 & Tier 4 \\
OpenAlex2  & 450.0 & 194{,}890 & 4.3 & Tier 4 \\
OpenAlex3  & 436.5 & 72{,}981 & 4.3 & Tier 4 \\
\bottomrule
\end{tabular}
\end{table}

\begin{table}[H]
\centering
\caption{Venue tier distribution by split.}
\label{tab:venue_distribution_app}
\setlength{\tabcolsep}{1pt}
\begin{tabular}{lccccc}
\toprule
\textbf{Dataset} & \textbf{Tier 1} & \textbf{Tier 2} & \textbf{Tier 3} & \textbf{Tier 4} & \textbf{Tier 5} \\
\midrule
Aminer1          & 2 (0.0\%)     & 813 (6.8\%)   & 3190 (26.6\%) & 7698 (64.1\%) & 297 (2.5\%) \\
Aminer2          & 8 (0.1\%)     & 782 (6.5\%)   & 3215 (26.8\%) & 7681 (64.0\%) & 311 (2.6\%) \\
Aminer3          & 4 (0.0\%)     & 823 (6.9\%)   & 3152 (26.3\%) & 7736 (64.5\%) & 285 (2.4\%) \\
OpenAlex1 & 2 (0.0\%)     & 1094 (9.1\%)  & 2528 (21.1\%) & 7602 (63.3\%) & 774 (6.5\%) \\
OpenAlex2 & 1 (0.0\%)     & 1060 (8.8\%)  & 2541 (21.2\%) & 7652 (63.8\%) & 745 (6.2\%) \\
OpenAlex3 & 2 (0.0\%)     & 1080 (9.0\%)  & 2524 (21.0\%) & 7605 (63.4\%) & 789 (6.6\%) \\
\bottomrule
\end{tabular}
\end{table}

\end{document}